\documentclass[prl,superscriptaddress,aps,10pt,twocolumn]{revtex4}
\usepackage{amsmath,amssymb,graphicx,enumerate,bbm,color}
\usepackage{amsthm}
\usepackage{hyperref}

\providecommand{\ket}[1]{|#1\rangle}

\providecommand{\ketbra}[2]{|#1\rangle\!\langle#2|}

\newcommand{\proj}[1]{|#1\rangle\!\langle#1|}
\newcommand{\one}{\mathbbm{1}}
\newcommand{\tr}{{\rm tr}}

\renewcommand{\H}{\mathcal{H}}
\newcommand{\be}{\begin{equation}}
\newcommand{\ee}{\end{equation}}
\newcommand{\ba}{\begin{eqnarray}}
\newcommand{\ea}{\end{eqnarray}}

\begin{document}

\title{Measurement-Device-Independent Entanglement Witnesses \\ for All Entangled Quantum States}

\author{Cyril Branciard}
\affiliation{Centre for Engineered Quantum Systems and School of Mathematics and Physics, The University of Queensland, St Lucia, QLD 4072, Australia}
\author{Denis Rosset}
\affiliation{Group of Applied Physics, University of Geneva, Chemin de Pinchat 22, CH-1211 Geneva 4, Switzerland}
\author{Yeong-Cherng Liang}
\affiliation{Group of Applied Physics, University of Geneva, Chemin de Pinchat 22, CH-1211 Geneva 4, Switzerland}
\author{Nicolas Gisin}
\affiliation{Group of Applied Physics, University of Geneva, Chemin de Pinchat 22, CH-1211 Geneva 4, Switzerland}

\date{\today}

\begin{abstract}
The problem of demonstrating entanglement is central to quantum information processing applications. Resorting to standard entanglement witnesses requires one to perfectly trust the implementation of the measurements to be performed on the entangled state, which may be an unjustified assumption. Inspired by the recent work of F. Buscemi [Phys. Rev. Lett. {\bf 108}, 200401 (2012)], we introduce the concept of Measurement-Device-Independent Entanglement Witnesses (MDI-EWs), which allow one to demonstrate entanglement of {\it all} entangled quantum states with untrusted measurement apparatuses. We show how to systematically obtain such MDI-EWs from standard entanglement witnesses. Our construction leads to MDI-EWs that are loss-tolerant, and can be implemented with current technology.
\end{abstract}

\maketitle

Quantum entanglement~\cite{RMP_entanglement} --- an essential feature of quantum theory, describing nonclassical correlations between quantum systems --- is the key resource that gives quantum information processing applications their advantage over classical computing~\cite{Jozsa08082003}. Its characterization and verification, both from a theoretical and a practical point of view, are therefore crucial problems in quantum information science.

Several criteria have been proposed to distinguish entangled quantum states from separable ones. A simple one is based on the concept of {\it Entanglement Witnesses}~\cite{horodeckis1996,terhal2000}: for any entangled state $\rho$, there exists a Hermitian operator $W$ such that $\tr[W \rho] < 0$, while $\tr[W \sigma] \geq 0$ for all separable states $\sigma$. Such an operator $W$ --- called an Entanglement Witness (EW) --- can thus be used to detect the entanglement of $\rho$. Experimentally testing an EW requires one to be able to estimate $\tr[W \rho]$; this is typically done by decomposing $W$ as a linear combination of product Hermitian operators, estimated independently from different local measurements on each subsystem of $\rho$.
A drawback of this entanglement verification technique using standard EWs is, however, that it requires a perfect implementation of the measurements, so as to faithfully reconstruct $\tr[W \rho]$. Imperfect measurements can indeed lead to an erroneous estimation of $\tr[W \rho]$, and possibly to the wrong conclusion about the presence of entanglement, even if $\rho$ is separable~\cite{SeevinckUffink_01,Bruss.PhysRevA.76.012312,SeevinckUffink_07,Moroder_2010,DIEWs,Rosset:2012:PRA}.

A way to get around this difficulty is to rely on the (loophole-free) violation of a Bell inequality~\cite{bell_book}. Indeed, within quantum theory this can only be obtained when one performs measurements on an entangled state~\cite{Werner:1989zz}. A violation therefore guarantees the presence of entanglement, independently of the measurements actually performed, of the functioning of any device used in the experiment, as well as of the dimension of the underlying shared quantum state. The price to pay when considering such Device-Independent Entanglement Witnesses (DI-EWs)~\cite{DIEWs} is that not all entangled states can be detected: there are indeed (mixed) entangled states that can only generate locally-causal correlations, which satisfy all Bell inequalities~\cite{Werner:1989zz,Barrett:2002gu} when measured one copy at a time. While this problem can be circumvented to some extent when partial knowledge of the system or devices is available (see, e.g., Ref.~\cite{Moroder:PRA:2012} and references therein), the possibility of witnessing all entangled states via these approaches remains unclear.

Generalizing the concept of so-called ``nonlocal games''~\cite{Cleve04}, Buscemi recently introduced Bell-like scenarios where instead of classical inputs specifying which measurements to be performed, the participating parties receive {\it quantum inputs}~\cite{Buscemi:2012all}.
Interestingly, he showed that all entangled states can give an advantage over separable states in such scenarios. While this suggests a way to certify the entanglement of all non-separable states, the proof presented in~\cite{Buscemi:2012all} does not, however, provide any explicit method to do this.

In this Letter we show that any standard entanglement witness can be used to derive an explicit criterion for witnessing entanglement in the aforementioned scenario with quantum inputs. Such criteria do not depend on the particular functioning of the measuring devices, and thus provide {\it Measurement-Device-Independent Entanglement Witnesses} (MDI-EWs), offering an interesting level of robustness against imperfect implementations. To illustrate the power of  these criteria, we provide explicit MDI-EWs that can be used to witness the entanglement of any 2-qubit entangled Werner state, as well as noisy 3-qubit Greenberger-Horne-Zeilinger (GHZ) states that exhibit genuine tripartite entanglement.
These MDI-EWs are tolerant to some common form of losses, and can be implemented with linear optics using current technology.

\paragraph{Measurement-Device-Independent EWs.---}
For simplicity we consider first a bipartite scenario, the extension to multipartite cases being discussed below.

In the so-called ``semiquantum nonlocal games''~\cite{Buscemi:2012all}, the two separated parties, Alice and Bob, receive some quantum states $\tau_s$ (for Alice) and $\omega_t$ (for Bob), and must output some values $a$ and $b$, respectively. The correlation between these values is characterized by the conditional probability distribution $P(a,b | \tau_s, \omega_t)$.

While not being allowed to communicate during the game (although they can communicate before to agree on a pre-established strategy --- see also Ref.~\cite{QI_scenario_NoSimul} where communication is allowed during the game), Alice and Bob are nonetheless allowed in this scenario to share some quantum state $\rho_{AB}$. If their state is entangled, then Alice and Bob can obtain --- by making appropriate joint measurements on their respective part of $\rho_{AB}$ and on their input states --- a correlation $P_{\rho}$ which cannot be explained without entanglement~\cite{Buscemi:2012all}. Indeed, convexity arguments allow one to prove that there must exist a linear combination of probabilities
\ba
I(P) \, = \, \sum_{s,t,a,b} \, \tilde\beta_{s,t,a,b} \ P(a,b | \tau_s, \omega_t) \label{eq_def_MDIEW}
\ea
such that $I(P_{\rho}) < 0$, while $I(P) \geq 0$ necessary holds if Alice and Bob only have access to separable quantum states, and possibly infinite shared randomness. 
The real coefficients $\tilde\beta_{s,t,a,b}$ in Eq.~\eqref{eq_def_MDIEW} correspond, up to a sign, to Buscemi's ``payoff function''~\cite{Buscemi:2012all}, multiplied by the probability distributions for $s$ and $t$. 

The fact that $I(P) \geq 0$ must hold when Alice and Bob do not share entanglement does not depend on the measurements they perform --- as long as these are fully described by quantum theory. Hence, if Alice and Bob obtain a value $I(P) < 0$, this guarantees in a {\it Measurement-Device-Independent} (MDI) manner (although not independently of the input states and of quantum theory) that they shared entanglement; in this sense, $I(P)$ from Eq.~\eqref{eq_def_MDIEW} defines a MDI-EW.
In particular, no assumption is made on the dimension of the shared state.
The requirement that with standard EWs the measurement implementations must be trusted is replaced here by the requirement that the input states must precisely be those specified in the assumptions. This is a natural assumption if one considers the framework of nonlocal games where an external referee wants to be convinced that two untrusted parties share entanglement; clearly, the referee trusts the input states that he/she chooses. Another conceivable situation would be a case where Alice and Bob want to verify their entanglement themselves, and therefore choose and generate their input states themselves; in that case their generating devices must be trusted. Arguably, it may however be more reasonable to trust a source device than a detection device (which is, by definition, open to its external, untrusted environment and may receive any kind of physical systems from it). This can indeed be the case in particular in the context of Quantum Key Distribution (QKD), where MDI-QKD --- with the similar assumptions that the state preparations are trusted, but not the measuring devices --- has been proposed~\cite{MDIQKD_Braunstein,MDIQKD_Lo,Charles:LocalBellTest} and recently implemented~\cite{MDIQKD_expmt}.
Note that in all cases, it is a crucial assumption when using MDI-EWs that the measuring devices must not have access {\it a priori} to the classical labels $s$, $t$ of the input states; there must in particular be no unwanted side-channels carrying this information to Alice and Bob's devices.

\paragraph{MDI-EWs from standard EWs.---}

We now explain how MDI-EWs can be derived from any standard EW.

Consider a bipartite entangled state $\rho_{AB}$ acting on some Hilbert space ${\cal H}_A \otimes {\cal H}_B$, with $\dim {\cal H}_A = d_A$ and $\dim {\cal H}_B = d_B$. Let $W$ be an EW detecting the entanglement of $\rho_{AB}$, i.e., a Hermitian operator on ${\cal H}_A \otimes {\cal H}_B$ such that $\tr[W \rho_{AB}] < 0$, while $\tr[W \sigma_{AB}] \geq 0$ for all separable states $\sigma_{AB} \in {\cal H}_A \otimes {\cal H}_B$. Because the set of density matrices spans the whole space of Hermitian operators, $W$ can be written in the (non-unique) form
\ba
W \, = \ \sum_{s,t} \ \beta_{s,t} \ \tau_s^\top \otimes \omega_t^\top \, , \label{eq_decomp_W}
\ea
with some real coefficients $\beta_{s,t}$, and where the operators $\tau_s^\top \in {\cal H}_A$ and $\omega_t^\top \in {\cal H}_B$ are density matrices. Defining $\tau_s$ and $\omega_t$ to be their transposes, with respect to some orthonormal bases $\{\ket{i}\}$ of ${\cal H}_A$ and $\{\ket{j}\}$ of ${\cal H}_B$, these are also density matrices.

We use the decomposition~\eqref{eq_decomp_W} to obtain a MDI-EW in the following way: Alice and Bob's inputs are the quantum states $\tau_s$ and $\omega_t$, respectively. They are asked to output a binary result, 0 or 1.
The expression
\ba
I(P) = \, \sum_{s,t} \ \beta_{s,t} \ P(1,1|\tau_s,\omega_t) \label{eq_MDIEW_from_EW}
\ea
is then of the form~\eqref{eq_def_MDIEW} [with $\tilde\beta_{s,t,1,1} = \beta_{s,t}$ and $\tilde\beta_{s,t,a,b} = 0$ for $(a,b) \neq (1,1)$], and takes non-negative values $I(P) \geq 0$ when Alice and Bob do not share entanglement, while sharing the entangled state $\rho_{AB}$ (and projecting their part of it together with their input state onto a maximally entangled state---see the proof below) allows them to get a correlation $P_\rho$ such that $I(P_{\rho}) < 0$. These properties precisely define a MDI-EW.

\begin{proof}

(i) Suppose Alice and Bob do not share entanglement; all they can do is then to share a separable state of the form $\sigma_{AB} = \sum_k p_k\ \sigma_{A}^k \otimes \sigma_{B}^k$, of any dimension, with $p_k \geq 0$ and $\sum_k p_k = 1$ (note that any shared randomness can be included in $\sigma_{AB}$), and measure their respective part of $\sigma_{AB}$ together with their input states. Writing $A_{1}$ and $B_{1}$ their Positive Operator-Valued Measure (POVM) elements~\cite{NielsenChuang} corresponding to the outcomes $1$, the probability that they both get this outcome is
\ba
P_\sigma(1,1|\tau_s,\omega_t) &=& \tr[(A_{1} \otimes B_{1}) \cdot (\tau_s \otimes \sigma_{AB} \otimes \omega_t)] \nonumber \\
&=& \sum_k \ p_k \ \tr[(A_{1}^k \!\otimes B_{1}^k) \cdot (\tau_s \otimes \omega_t)] \qquad
\label{Eq:Prob}
\ea
with $A_{1}^k = \tr_A[A_{1} \cdot (\one \otimes \sigma_{A}^k)]$ and $B_{1}^k = \tr_B[B_{1} \cdot (\sigma_{B}^k \otimes \one)]$ being effective POVM elements acting on the state space of $\tau_s$ and $\omega_t$, and where $\tr_A$ and $\tr_B$ denote the partial traces over the systems of $\sigma_{A}^k$ and $\sigma_{B}^k$.
From Eqs.~\eqref{eq_MDIEW_from_EW} and~\eqref{eq_decomp_W} and by linearity, one finds
\ba
I(P_\sigma) &=& \sum_{s,t} \ \beta_{s,t} \ \sum_k \ p_k \ \tr[(A_{1}^k \otimes B_{1}^k) \cdot (\tau_s \otimes \omega_t)] \nonumber \\
&=& \sum_k \ p_k \ \tr[(A_{1}^k \otimes B_{1}^k) \cdot W^{\top}] \, .
\label{Eq:CalcI}
\ea
Since $W$ is an an entanglement witness and since all operators $A_{_{1}}^{k\top}$ and $B_{1}^{k\top}$ are positive Hermitian operators, one has $\tr[(A_{1}^k \otimes B_{1}^k) \cdot W^{\top}] = \tr[W \cdot (A_{1}^{k\top} \otimes B_{1}^{k\top})] \geq 0$ for all $k$. This proves that $I(P_\sigma) \geq 0$ for any separable state $\sigma_{AB}$.

\medskip

(ii) Suppose now that Alice and Bob share the state $\rho_{AB}$. We define Alice and Bob's measurements to be the joint projections of their input state and their part of $\rho_{AB}$ onto the maximally entangled states $\ket{\Phi^+_{AA}} = \frac{1}{\sqrt{d_A}} \sum_{i=1}^{d_A} \ket{i} \otimes \ket{i}$ and $\ket{\Phi^+_{BB}} = \frac{1}{\sqrt{d_B}} \sum_{j=1}^{d_B} \ket{j} \otimes \ket{j}$; the outcome $1$ indicates a successful projection. We thus obtain
\ba
&& \hspace{-5mm} P_\rho(1,1|\tau_s,\omega_t) \nonumber \\
&& = \tr \big[ (\ketbra{\Phi^+_{\!AA}}{\Phi^+_{\!AA}} \! \otimes \ketbra{\Phi^+_{\!BB}}{\Phi^+_{\!BB}}) \cdot (\tau_s \otimes \rho_{AB} \otimes \omega_t) \big] \nonumber \\
&& = \tr \big[ (\tau_s^\top \otimes \omega_t^\top) \cdot \rho_{AB} \big] / (d_A d_B) \label{Eq:CalcViolation1}
\ea
and
\ba
I(P_\rho) &=& \sum_{s,t} \ \beta_{s,t} \ \tr \big[ (\tau_s^\top \otimes \omega_t^\top) \cdot \rho_{AB} \big] / (d_A d_B) \nonumber \\
&=& \tr \big[ W \rho_{AB} \big] / (d_A d_B) \ < \ 0, \label{Eq:CalcViolation2}
\ea
which concludes the proof.
\end{proof}

Note that our construction above gives a more direct proof --- based simply on the already established existence of a standard EW for any entangled state~\cite{horodeckis1996} --- that all entangled states, including multipartite {\it non-fully-separable states}~\cite{RMP_entanglement,Guhne20091}, can be ``witnessed'' in a Bell-like scenario with quantum inputs~\cite{Buscemi:2012all}. An interesting feature of our construction is that it does not require Alice and Bob to perform a full (generalized) Bell measurement, as considered in Ref.~\cite{Buscemi:2012all}, but simply a projection onto a single maximally entangled state.
We emphasize also that the proof that the separable bound is 0 allows for any possible measurement by Alice and Bob, together also with any pre-measurement quantum operation, including losses --- whether they are state-dependent (e.g. polarization-dependent), on the input states and/or on the shared state. Losses therefore cannot lower this separable bound, and our MDI-EWs are not prone to any detection loophole, contrary to standard EWs~\cite{Bruss.PhysRevA.76.012312} (and to Bell inequalities when one postselects on the detected events~\cite{pearle1970hve}).
On the other hand, if an entangled state $\rho$ gives $I(P_\rho) < 0$ without losses, and if the effect of losses is simply to reduce all probabilities $P(1,1|\tau_s,\omega_t)$ by the same multiplicative factor, then although the value of $I(P_\rho)$ will get closer to 0, it will remain negative with such losses; our MDI-EWs are thus resistant to this typical kind of losses.

\paragraph{MDI-EWs from standard EWs for genuine multipartite entanglement.---}

The construction given above can also be applied to entanglement witnesses that detect {\it genuine multipartite entanglement}~\cite{Guhne20091}. For simplicity, we shall concentrate here on the tripartite scenario, the generalization to an arbitrary number of parties being straightforward. Let us start by recalling that a tripartite state $\sigma_{\mathrm{bs}}$ shared among Alice, Bob and Charlie is said to be {\it biseparable} if it can be written in the form~\cite{Acin.PhysRevLett.87.040401,Guhne20091}:
\begin{equation}\label{Eq:Biseparable}
	\sigma_{\mathrm{bs}}=\sum_k \sigma_{AB}^k\otimes \sigma_{C}^k+\sum_k \sigma_{AC}^k\otimes \sigma_{B}^k+\sum_k \sigma_{BC}^k\otimes \sigma_{A}^k\,,
\end{equation}
where we have included the weight of each individual state of the mixture in its normalization; a state that cannot be written as above is said to be {\it genuinely tripartite entangled}.

Consider such a genuinely tripartite entangled state $\rho_{ABC}$ acting on $\H_A\otimes\H_B\otimes\H_C$, and a witness $W$ detecting its genuine tripartite entanglement --- i.e., a Hermitian operator such that $\tr[W\, \rho_{ABC}] < 0$, while $\tr[W \sigma_{\mathrm{bs}}] \geq 0$ for all biseparable states $ \sigma_{\mathrm{bs}}$ acting on $\H_A\otimes\H_B\otimes\H_C$~\cite{footnote_W_genuine_entgmt}. Similarly to Eq.~\eqref{eq_decomp_W}, let us decompose $W$ as:
\begin{equation}
	W \, = \ \sum_{s,t,u} \ \beta_{s,t,u} \ \tau_s^\top \otimes \omega_t^\top\otimes \gamma_u^\top \label{eq_decomp_W_GTE},
\end{equation}
where $\tau_s^\top$, $\omega_t^\top$, and $\gamma_u^\top$ are density matrices acting on $\H_A$, $\H_B$, and $\H_C$ while $\beta_{s,t,u}$ are real expansion coefficients.

In a similar way as in the bipartite case, one can show that using the input states $\tau_s$ for Alice, $\omega_t$ for Bob and $\gamma_u$ for Charlie, and letting them perform binary-outcome measurements, the inequality
\begin{equation}
	I(P) = \, \sum_{s,t,u} \ \beta_{s,t,u} \ P(1,1,1|\tau_s,\omega_t,\gamma_u)\ge 0 \label{eq_MDIEW_from_EW_GTE}
\end{equation}
holds true for all probability distributions $P(1,1,1|\tau_s,\omega_t,\gamma_u)$ obtainable from biseparable states $\sigma_{\mathrm{bs}}$, but can be violated by some probability distributions derived from $\rho_{ABC}$. The proof follows closely that for the bipartite case given previously. We only sketch below the less trivial part for the benefit of the readers.

Let us concentrate for now on the first sum in Eq.~\eqref{Eq:Biseparable}. In analogy to Eq.~\eqref{Eq:Prob}, we get (with some abuse of notations in the ordering of the tensor products)
\begin{align}
	P_\sigma^{\mbox{\tiny AB$|$C}}&(1,1,1|\tau_s,\omega_t,\gamma_u)\nonumber\\
	= &\sum_k  \ \tr[(A_{1}\!\otimes B_{1} \!\otimes C_{1}) \cdot (\tau_s \otimes \omega_t\otimes \gamma_u\otimes \sigma_{AB}^k\otimes \sigma_C^k)] \nonumber\\
	= &\sum_k   \ \tr[(AB_{1}^k \!\otimes C_{1}^k) \cdot (\tau_s \otimes \omega_t\otimes \gamma_u)] 
\end{align}
where $A_{1}$, $B_{1}$, $C_{1}$ are respectively the local POVM elements of Alice, Bob and Charlie corresponding to their outcomes $1$, while $AB_{1}^k = \tr_{AB}[A_{1}\otimes B_{1}  \cdot (\one \otimes \sigma_{AB}^k)]$, $C_{1}^k = \tr_C[C_{1} \cdot (\one \otimes \sigma_{C}^k)]$, and the partial traces $\tr_{AB}$, $\tr_C$ are performed respectively over the state space of $\sigma_{AB}^k$ and $\sigma_C^k$. One can interpret the operators $AB_{1}^k$ and $C_{1}^k$ as effective POVM elements acting on the state space of $\tau_s\otimes\omega_t$ and $\gamma_u$; note that these (as well as their transposes) are non-negative Hermitian operators. As a result, as with the calculation detailed in Eq.~\eqref{Eq:CalcI}, the contribution of $P_\sigma^{\mbox{\tiny AB$|$C}}$ towards $I(P)$ in Eq.~\eqref{eq_MDIEW_from_EW_GTE} is non-negative; likewise for the other terms from Eq.~\eqref{Eq:Biseparable}. All in all, we  thus see that inequality~\eqref{eq_MDIEW_from_EW_GTE} holds true for all correlations obtainable from any tripartite biseparable state $\sigma_{\mathrm{bs}}$. Lastly, to see that inequality~\eqref{eq_MDIEW_from_EW_GTE} can be violated by $\rho_{ABC}$, one can proceed with a similar calculation as that detailed in Eqs.~\eqref{Eq:CalcViolation1} and \eqref{Eq:CalcViolation2}, letting Alice, Bob and Charlie perform projections onto maximally entangled states.

We note that such MDI-EWs for multipartite entanglement are also resistant to the typical kind of losses mentioned before, although in that case, the values of $I(P_\rho)$ will approach the separable bound 0 faster as the number of qubits gets larger.

\paragraph{Some explicit examples of  MDI-EWs.---}

In order to illustrate our technique for constructing MDI-EWs, let us first consider the 2-qubit Werner state~\cite{Werner:1989zz}
\ba
\rho_{AB}^v \ = \ v \, \ketbra{\Psi^-}{\Psi^-} + (1-v) \, \one/4 \label{def_Werner_state}
\ea
with $v \in [0,1]$ and where $\ket{\Psi^-} = \frac{\ket{01} - \ket{10}}{\sqrt{2}}$ is the singlet state. $\rho_{AB}^v$ is entangled if and only if $v > 1/3$~\cite{Werner:1989zz,Peres:1996separability}, which can be detected with the EW~\cite{Toth.PhysRevLett.94.060501,Guhne20091}
\ba
W = \frac{1}{2} \one - \ketbra{\Psi^-}{\Psi^-},
\ea
such that $\tr [W \rho_{AB}^v] = \frac{1-3v}{4} < 0$ for $v > 1/3$, while $\tr [W \sigma_{AB}] \geq 0$ for all separable 2-qubit states $\sigma_{AB}$.

$W$ can for instance be decomposed as in Eq.~\eqref{eq_decomp_W} with:
\ba
\beta_{s,t} = \frac{5}{8} \ \text{ if } s = t, \quad \beta_{s,t} = -\frac{1}{8} \ \text{ if } s \neq t,
\ea
for $s,t = 0,\ldots,3$, and
\ba\label{Eq:InputState}
\tau_s = \sigma_s \frac{\mathbbm{1} \!+ \vec{n} \cdot \vec{\sigma}}{2} \sigma_s, \quad \omega_t = \sigma_t \frac{\mathbbm{1} \!+ \vec{n} \cdot \vec{\sigma}}{2} \sigma_t,
\ea
where $\vec{n} = (1,1,1)/\sqrt{3}$, $\vec{\sigma} = (\sigma_1,\sigma_2,\sigma_3)$ is the vector of Pauli matrices, and where we write $\sigma_0 = \mathbbm{1}$ for convenience. The MDI-EW we obtain using our construction can thus explicitly be written as
\ba
I(P) = \, \frac{5}{8} \sum_{s=t} P(1,1|\tau_s,\omega_t) - \frac{1}{8} \sum_{s \neq t} P(1,1|\tau_s,\omega_t), \quad
\ea
with Alice and Bob's 4 pure states $\tau_s,\omega_t$ corresponding to the four vertices of a regular tetrahedron on the Bloch sphere.
By construction, $I(P) \geq 0$ must hold if Alice and Bob do not share entanglement, while they can obtain $I(P_{\rho^v}) = \frac{1-3v}{16} < 0$ if they share the Werner state $\rho_{AB}^v$ of Eq.~\eqref{def_Werner_state} with $v > \frac{1}{3}$ and perform joint projective measurements on their part of $\rho_{AB}^v$ and their input states, onto the maximally entangled states $\ket{\Phi^+} = \frac{\ket{00} + \ket{11}}{\sqrt{2}}$ (or $\ket{\Psi^-}$, equivalently).

Note that another possible decomposition for $W$ is the following: defining $s = (s_1,s_2)$ and $t = (t_1,t_2)$, with $s_1,t_1 = 0, 1$ and $s_2,t_2 = 1,\ldots,3$, Eq.~\eqref{eq_decomp_W} holds with
\ba
\beta_{s,t} \ = \ \delta_{s_2,t_2} \, \frac{3 \, \delta_{s_1,t_1} - 1}{6} ,
\ea
where $\delta_{i,j}$ is the Kronecker delta, and with
\ba
\tau_s = \frac{\mathbbm{1} \!+ (-1)^{s_1} \sigma_{s_2}}{2}, \quad \omega_t = \frac{\mathbbm{1} \!+ (-1)^{t_1} \sigma_{t_2}}{2}. \label{Eq:InputState2}
\ea
This defines another MDI-EW, now with the input states $\ket{\pm x}, \ket{\pm y}, \ket{\pm z}$, the respective eigenstates of the three Pauli matrices.

As another example, let us consider the family of noisy $3$-partite GHZ states
\begin{equation}
	\rho^v_{GHZ} \ = \ v \, \proj{\text{GHZ}} + (1-v) \, \one/8,
\end{equation}
where $\ket{\text{GHZ}} = \tfrac{\ket{000}+\ket{111}}{\sqrt{2}}$. This state is genuinely tripartite entangled if and only if $v>3/7$~\cite{Guhne:2010:NJP}, which can be demonstrated using the following witness for genuine tripartite entanglement~\cite{Toth.PhysRevLett.94.060501,Guhne20091}:
\begin{equation}
  W_\text{GHZ} = \frac{1}{2}\one - \proj{\text{GHZ}}.
\end{equation}
A possible way to decompose  $W_\text{GHZ}$  in the form of Eq.~\eqref{eq_decomp_W_GTE}  is to make use of the states defined in Eq.~\eqref{Eq:InputState} (with $\gamma_u$ defined analogously), together with the coefficients
\ba
	\beta_{s,t,u} &=& \frac{3}{32} \, (-1)^{\lfloor\!\frac{s{-}1}{2}\!\rfloor \lfloor\!\frac{t{-}1}{2}\!\rfloor + \lfloor\!\frac{s{-}1}{2}\!\rfloor \lfloor\!\frac{u{-}1}{2}\!\rfloor + \lfloor\!\frac{t{-}1}{2}\!\rfloor \lfloor\!\frac{u{-}1}{2}\!\rfloor + 1} \nonumber \\[-1mm]
	&& \hspace{7mm} \times \big[ (-1)^{\lfloor\!\frac{s{-}1}{2}\!\rfloor + \lfloor\!\frac{t{-}1}{2}\!\rfloor + \lfloor\!\frac{u{-}1}{2}\!\rfloor} + (-1)^{s+t+u}\sqrt{3} \big] . \nonumber
\ea
These define a MDI-EW as in~\eqref{eq_MDIEW_from_EW_GTE}, allowing one to certify the genuine tripartite entanglement of $\rho^v_{GHZ}$ for all $v>3/7$ (by again letting Alice, Bob and Charlie perform projections onto $\ket{\Phi^+}$).

\paragraph{Conclusion.---}

Inspired by Buscemi's ``semiquantum nonlocal games'', we have introduced the concept of Measurement-Device-Independent Entanglement Witnesses.
These can certify the entanglement of all entangled states, however weakly entangled they are --- which cannot be done in a fully device-independent manner (at least, when measured one copy at a time~\cite{Werner:1989zz,Barrett:2002gu}; see however~\cite{Cavalcanti:arXiv2012}).
We have shown that explicit, fairly loss-tolerant MDI-EWs can be systematically obtained from {\it any} standard entanglement witness (and actually, from any particular decomposition thereof) --- including those for witnessing genuine multipartite entanglement --- in a scenario where each party receives quantum input states and projects them, together with their part of an entangled state, onto a maximally entangled state.

For multi-qubit entangled states, as considered in the explicit examples above, the MDI-EWs obtained with our construction thus only involve projections onto 2-qubit Bell states, which can be implemented with linear optics by simply letting photons interfere on a beam-splitter~\cite{Weinfurter1994}. Together with their loss-tolerance, this makes our MDI-EWs quite suitable for experimental tests, which are readily feasible with current technology. 
An experimental challenge will nevertheless be to faithfully prepare the input states with a very good precision and no side-channel (which may otherwise transmit their classical labels), and possibly to adapt our MDI-EWs to account for some small, experimentally inescapable imprecision in the input state preparation (e.g. by calculating new separable bounds, as can be done to account for measurement misalignments in the case of standard entanglement witnesses~\cite{Rosset:2012:PRA}).

With our approach, the problem of finding MDI-EWs reduces to the problem of finding standard EWs, for which a number of techniques are known (see, e.g.,~\cite{Doherty:PRA:2004} and references therein). Note that our construction only generates MDI-EWs of the special form of Eq.~\eqref{eq_MDIEW_from_EW} (and that reciprocally, any MDI-EW of the form~\eqref{eq_MDIEW_from_EW} which can be tested by Alice and Bob performing projections onto maximally entangled states gives rise through Eq.~\eqref{eq_decomp_W} to a standard EW).
More general forms of MDI-EWs exist, as in~\eqref{eq_def_MDIEW}, which may be useful, for instance, if the projection onto a maximally entangled state happened to be infeasible. Other techniques will however be required to generate more general MDI-EWs. It may be interesting to investigate whether more general MDI-EWs can have other practical advantages, e.g. with respect to the robustness to imperfect state preparation, or for possible applications like MDI-QKD based on MDI-EWs; the ideas developed here indeed suggest the possibility for new protocols for MDI-QKD, different from those of~\cite{MDIQKD_Braunstein,MDIQKD_Lo,Charles:LocalBellTest,MDIQKD_expmt}. Likewise, another interesting question is whether using more quantum inputs (as in~\eqref{Eq:InputState2} compared to~\eqref{Eq:InputState}) offers some more robustness, or some advantage for MDI-QKD. These are left as open problems.

\paragraph{Acknowledgments.---} We acknowledge useful discussions with Francesco Buscemi. This work was supported by a UQ Postdoctoral Research Fellowship, by the Swiss NCCR ``Quantum Science and Technology", the CHIST-ERA DIQIP, and the European ERC-AG QORE.

\paragraph{Note added.---} While finalizing our manuscript, we became aware of the recent work of Cavalcanti {\it et al.}~\cite{CavalcantiHallWiseman}, which interprets (and extends) Buscemi's results~\cite{Buscemi:2012all} in terms of trust in the entanglement verification procedure. Our results answer in particular some of the questions mentioned in the conclusion of that paper.

$ $


\begin{thebibliography}{33}
\bibitem[{foo()}]{footnote_W_genuine_entgmt}
\bibinfo{note}{Because the set of biseparable states is convex, note that for
  any genuinely tripartite entangled state $\rho$, there always exists a
  witness $W$ satisfying the desired properties.}
\end{thebibliography}

\end{document}